\documentstyle[12pt]{article}
\begin{document}
\title{The high-temperature specific heat exponent of the 3-d Ising model}
\author{
A.J. Guttmann\\Department of Mathematics,
The University of Melbourne\\ Parkville, Vic. 3052, Australia\\
and I.G. Enting\\CSIRO Division of Atmospheric Research,\\Private Bag No.1\\
Mordialloc, Vic. 3195, Australia}
\date{}
\maketitle
\bibliographystyle{unsrt}
\begin{abstract}
We have extended the high-temperature susceptibility series of the
three-dimensional
spin-$\frac{1}{2}$ Ising model to $O(v^{26})$. Analysis of the new series gives
$\alpha = 0.101 \pm 0.004$.
\end{abstract}
In an earlier paper \cite{ge1} we gave series to order $v^{22}$ for the
high-temperature expansion of the zero-field partition function of the
3-dimensional
Ising model. More precisely, we gave the coefficients $a_n, \mbox{  } n=0,22$,
defined by \[Z =2[cosh(J/kT)]^3\Phi(v), \mbox{  with  } \Phi(v)=\sum_{n}a_n
v^n.\]
The series were obtained by the finite-lattice method. One
difficulty with the finite-lattice method for this problem is its voracious
appetite for computer memory. Our earlier computation in fact calculated
the series to two further terms - to order  $v^{26}$ - but due to addressing
limitations, we were unable to retain the intermediate information. This
particular calculation requires $2.08 GB$ of memory, and we were unable to
address more than $2 GB$, due to operating system limitations. We have now
been able to re-run our program under a different operating system that permits
us to address this large address space.

The program was run on an IBM 3090/400J with $500 MB$ of memory and $2 GB$ of
extended storage - a slower type of memory. The use of the MVS operating system
allowed the large address space to be used. Even so, 2-byte integers were
used, and the program run twice {\em modulo} two different primes. The results
were combined using the Chinese Remainder Theorem, and provided the least
significant
digits of the new coefficients, the most significant digits were obtained by
differential approximants. The final results were then compared by running with
a third prime. Each run took 150 hours.

As a result, we have obtained two further non-zero terms (the partition
function
being an even function has vanishing odd-order coefficients). We have also
obtained the 6 most significant digits of the $O(v^{28})$ coefficient, by
the method of differential approximants. In our earlier paper we obtained
the coefficient of the $O(v^{24})$ coefficient by this method, and claimed the
coefficient to be $a_{24} = 27337 * 10^7$. The present calculation gives the
coefficient
as $a_{24}=273374177222$, verifying our prediction. The subsequent coefficients
are found to be $a_{26}= 4539862959852$ and $a_{28} = 7474452 * 10^7 \pm 5 *
10^7$,
where the last coefficient is obtained by differential approximants. (The
approximate coefficient was not used in the subsequent analysis,
as differential approximants
require more accurate coefficients. It is nevertheless useful for ratio type
methods
of analysis).

As we were completing this work we received a preprint \cite{bc} in which a
variant of the finite-lattice method, using helical boundary conditions was
used to obtain one further coefficient than we had previously obtained. This
work also confirmed our predicted coefficient, and agrees with our exact
coefficient. (Note that they give the free-energy series and we give
the partition function series). They also predicted $a_{26}$, and our exact
coefficient confirms
their predicted value.

We have analysed the new series by several methods. The series is now, for the
first time, sufficiently long that the method of differential approximants can
be used with some confidence. For our initial analysis, we used unbiased
approximants,
but for maximum precision we used biased approximants. This requires a
knowledge of
the critical temperature which has been accurately estimated from
the more readily analysed high-temperature susceptibility series, as well as
from
a variety of Monte-Carlo estimates. The series estimates are
reviewed in \cite{g1}, and we use the best estimate given there, $v_c =
0.218093$,
which is in good agreement with the most recent, high-precision Monte-Carlo
estimate,
\cite{l} of $v_c = 0.2180992 \pm 0.0000026$.

Our method of analysis is fully described in \cite{g2},
and provides a weighted mean of critical exponent estimates from inhomogeneous
first-
and second-order differential approximants, with one estimate obtained for each
order
of the series. Our analysis was carried out on the coefficients of the
partition
function itself. Our unbiased estimates are \[v_c^2 = 0.04756 \pm 0.00003
\mbox{  and  }
2-\alpha = 1.905 \pm 0.016 \mbox{  with  } K=1\]
\[v_c^2 = 0.04756 \pm 0.00002 \mbox{  and  }
2-\alpha = 1.897 \pm 0.012 \mbox{  with  } K=2.\]
In the above, $K=1,2$ refers to first- and second-order differential
approximants
respectively. The unbiased estimates are seen to be in excellent agreement with
the sussceptibility series estimate $v_c^2 = 0.0475646$, while
 an estimate of $\alpha = 0.10 \pm 0.01$ can be made. A biased analysis yields
the following:

\[2-\alpha = 1.899 \pm 0.004,\mbox{   } K=1  \mbox{  and  }
2-\alpha = 1.900 \pm 0.006, \mbox{   } K=2\]
Thus we find, from this analysis, $\alpha = 0.101 \pm 0.004$.
This is substantially more precise than our earlier analysis, using two fewer
series coefficients,
of $\alpha = 0.104 \pm 0.018$. It is consistent with the analysis of \cite{bc}
who
find $\alpha = 0.104 \pm 0.004$, though as can be seen we favour a rather lower
value.
Note that second-order differential approximants implicitly take
correction-to-scaling
terms into account. The agreement between first- and second-order differential
approximants suggests that correction-to-scaling exponents are weak. A
subsequent
analysis provides numerical confirmation of this.

Ratio techniques can also be used with this series.
We have analysed the free-energy series by
a variety of extrapolation methods, based on the observation that if the
free-energy,
$\Psi/kT \sim A(1 - v^2/v_c^2)^{2-\alpha}$, then the ratio of successive
coefficients
in the series expansion of $\Psi/kT$ behaves like
$\frac{1}{v_c^2}(1 + \frac{\alpha-3}{n})$,
with higher order corrections from correction-to-scaling exponents, as well as
corrections
due to analytic terms. In any event, the sequence of ratios can obviously be
re-arranged
to give a sequence that will converge to $\alpha$. Neville extrapolation (which
takes
into account only analytic correction terms), gives $\alpha = 0.103 \pm 0.006$.
Other
extrapolation methods, such as Levin's u-transform and Brezinski's
$\theta$-algorithm
are less accurate, allowing only the estimate $\alpha = 0.10 \pm 0.03$.

In our previous analysis, we also studied the amplitude of the
``correction-to-scaling''
term, $a_\theta$, where the specific heat is defined to have the scaling form
$C \sim A|t|^{-\alpha}[1 + a_\theta|t|^{\theta} + a_1|t| + \cdots]$, where
$t=(T-T_c)/T_c$
and $\theta \approx 0.52$ \cite{ni}. In \cite{lf} it was argued that
$a_\theta$ should be negative,
and our earlier analysis \cite{ge1} seemed to confirm this, in that we
found $a_\theta \approx -0.04$.
This can be seen from the behaviour of the ratios of successive coefficients,
as follows:
We first write $C(v) = \sum c_n v^{2n}$, as the expansion we obtain
is in terms of the usual high-temperature variable $v=$tanh$(J/kT)$.
Note that, to leading
order, $t=(T-T_c)/T_c=B(v-v_c)/v_c$, where B is a positive constant.
It therefore follows that the correction-to-scaling amplitude of the specific
heat
series expanded in
the variable $v^2$ should also be of negative sign. Writing \[C(v)= \sum c_n
v^{2n} = A(1 -
 v^2/v_c^2)
^{-\alpha}(1 + b(1-v^2/v_c^2)^\theta +\cdots),\] it follows that
\[c_n = \frac{A\Gamma(\alpha+n)}{\Gamma(\alpha)\Gamma(n+1)v_c^{2n}}[1 +
\frac{b\Gamma(\alpha)\Gamma(\alpha+n-\theta)}{\Gamma(\alpha-\theta)\Gamma(\alpha+n)}
+ \cdots\] hence
\[\frac{c_n}{c_{n-1}} = \frac{1}{v_c^2}[1 + \frac{\alpha-1}{n} -
\frac{b\Gamma(\alpha)\theta}{\Gamma(\alpha-\theta)n^{\theta+1}} +
O(\frac{1}{n^2})].\]
Taking $\alpha \approx 0.1$ and $\theta \approx 0.5$, it follows that the above
equation
can be rewritten as
\[\frac{c_n}{c_{n-1}} = \frac{1}{v_c^2}[1 + \frac{\alpha-1}{n} +
\frac{1.28..b}{n^{\theta+1}} + O(\frac{1}{n^2})].\]
Hence we find that
\[(\frac{c_n}{c_{n-1}}v_c^2 - 1)n + 1 \sim \alpha + \frac{1.28..b}{n^\theta} +
O(\frac{1}{n
}).\]
This means that if $b < 0$, estimators of $\alpha$, given by
the l.h.s. of the above equation, should approach $\alpha$
{\em from below}. In fact we find the approach to be from above,
but a simple $n$-shift of 1
makes the approach change to an approach from below!  Even an analysis taking
into account
the analytic correction term does not alter this behaviour. To be more precise,
we have
repeated the above analysis with an additional analytic correction-to-scaling
term present,
and found that the numerical value of $b$ changes sign with an $n$-shift of
just $1$.
In all cases, the estimate of $b$ is numerically rather small, and we conclude
that this
analysis is not sensitive enough to distinguish $b$ from zero. A similar
conclusion,
based on a somewhat different analysis, was obtained in \cite{bc}.

Our estimate of $\alpha$ is rather lower than the field-theory estimate
\cite{lz} of
$\alpha = 0.110 \pm 0.0045$, but the field-theory and series estimates are both
(separately)
consistent with the hyperscaling relation $d\nu = 2-\alpha$. Our best series
estimate
of $\nu = 0.632^{+0.002}_{-0.003}$ implies $\alpha = 0.104^{+0.006}_{-0.009}$,
while
the best field-theory estimate \cite{ni} is $\nu = 0.630$, which implies
$\alpha = 0.110$,
a value at the centre of the field-theory estimates.

We summarise the various estimates of $\alpha$ in table \ref{t1}.

\begin{table}
{\hfill
\begin{tabular}{||l|l|l||}
\hline
$\alpha$ estimate & Method &  Reference \\
\hline
$0.101(4)$ & Series & This work \\
$0.104(4)$ & Series & \cite{bc} \\
$0.1100(45)$ & Field theory & \cite{lz} \\
$0.104_{-0.009}^{+0.006}$ & Series + hyperscaling & \cite{g2} \\
$0.110$ & Field theory  + hyperscaling & \cite{ni} \\
\hline
\end{tabular}
\hfill}
\caption{Summary of $\alpha$ estimates}
\label{t1}
\end{table}
\section*{Acknowledgements}
We wish to thank Mr. Bob Hill of the Australian Communications and Computing
Institute
for the provision of the computing facilities, and for assistance in running
this
large job. Financial support from the Australian Research Council is also
gratefully acknowledged.

\end{document}